\documentclass{aa} %

\usepackage{graphicx}
\usepackage{txfonts}
\newcommand{\Ms}{$\textrm{M}_{\odot}$}

\newcommand{\kms}{$\textrm{km~s$^{-1}$}$}
\newcommand{\Ha}{H$\alpha$}

\begin{document} 

\title{Star formation in outer rings of S0 galaxies. V.}

   \subtitle{UGC 4599 -- an S0 with gas probably accreted from a filament.}

   \author{O. Sil'chenko
          \inst{1}
          \and
          A. Moiseev
          \inst{2,1}
          \and
          D. Oparin
          \inst{2}
          \and
          J. E. Beckman
          \inst{3,4}
          \and
          J. Font
          \inst{3,4,5}
          }

   \institute{Sternberg Astronomical Institute of the Lomonosov Moscow
             State University, University av. 13, 119234 Russia\\
             \email{olga@sai.msu.su}
          \and
             Special Astrophysical Observatory
             of the Russian Academy of Sciences,
             Nizhnij Arkhyz, 369167 Russia\\
              \email{moisav@gmail.com,doparin@mail.ru}
        \and
        Instituto de Astrof\'isica de Canarias, 38205 La Laguna, Tenerife, Spain
        \and
        Departamento de Astrof\'isica, Universidad de La Laguna, Tenerife, Spain
        \and
        Observatorio Gemini Sur, NOIRLAB, La Serena, Chile
             }

   \date{Received  .., 2022; accepted .., 2022}

 
  \abstract
   {}
   {Though S0 galaxies are usually thought to be `red and dead', they often
   demonstrate weak star formation organised in ring structures and located in their
outer disks. We try to clarify the nature of this phenomenon and its difference from star
   formation in spiral galaxies. The moderate-luminosity nearby S0 galaxy, UGC 4599, is studied here.}
 {By applying long-slit spectroscopy at the Russian 6m telescope, we have measured stellar
kinematics for the main body of the galaxy and strong emission-line flux ratios in the ring. 
After inspecting the gas excitation in the ring using line ratio diagrams  and having shown that it is ionised by young stars, we have determined the gas oxygen abundance by
using conventional strong-line calibration methods. We have inspected the gas kinematics in the ring with Fabry-Perot interferometre data obtained at the William Herschel Telescope. The pattern and properties of the brightest
star formation regions are studied with the tunable filter MaNGaL at the 2.5m telescope of the Caucasian Mountain Observatory
of the Sternberg Astronomical Institute (CMO SAI MSU).}
{The gas metallicity in the ring is certainly subsolar, {\bf [O/H]$=-0.4 \pm 0.1$~dex}, that is different from the majority
of the outer starforming rings in S0s studied by us which have typically nearly solar metallicity. The total
stellar component of the galaxy which is old in the center is less massive than its extended gaseous disk. We conclude 
that probably the ring and the outer disk of UGC~4599 are a result of gas accretion from a cosmological filament.}
{}
\maketitle

\section{Introduction}

The morphological type S0 was initially introduced as star-formation free disk galaxies \citep{hubble},
while outer rings were early recognised as common attributes of S0 galaxies \citep{devauc59}.
Later a significant amount of cold gas has been found in many S0 galaxies, and in half of gas-rich S0s the gas feeds
star formation organised in ring structures \citep{pogge_esk93}. Moreover more than the half of outer
{\it stellar} rings in S0s are bright in ultraviolet (UV) so betraying recent star formation on a timescale of a few hundred Myr \citep{kostuk15}.
In the frame of the current paradigm according to which evolution of disk galaxies is driven by persistent accretion of
outer cold gas, this situation with S0s is quite understandable because in sparse environments the S0s may suffer
the same outer gas accretion as spirals, with a possibility of star formation in accreted gas. However, the
source of outer gas accretion remains still unknown. Cosmologists are sure that this source is provided
by large-scale Universe structure, in particular, by large-scale filaments filled by dark matter and primordial
gas \citep[e.g.]{lcdm}. The observations reveal nearly solar metallicity in the outer starforming rings of
S0 galaxies favouring rather gas-rich satellite merging \citep{s0_fp,u5936}. In this Letter we examine an outer
starforming ring in UGC~4599, a moderate-luminosity nearby S0 galaxy, which has indeed rather low metallicity and
can therefore be fed by primordial gas from cosmological filaments.

\begin{figure*}
\centering
	\includegraphics[width=18cm]{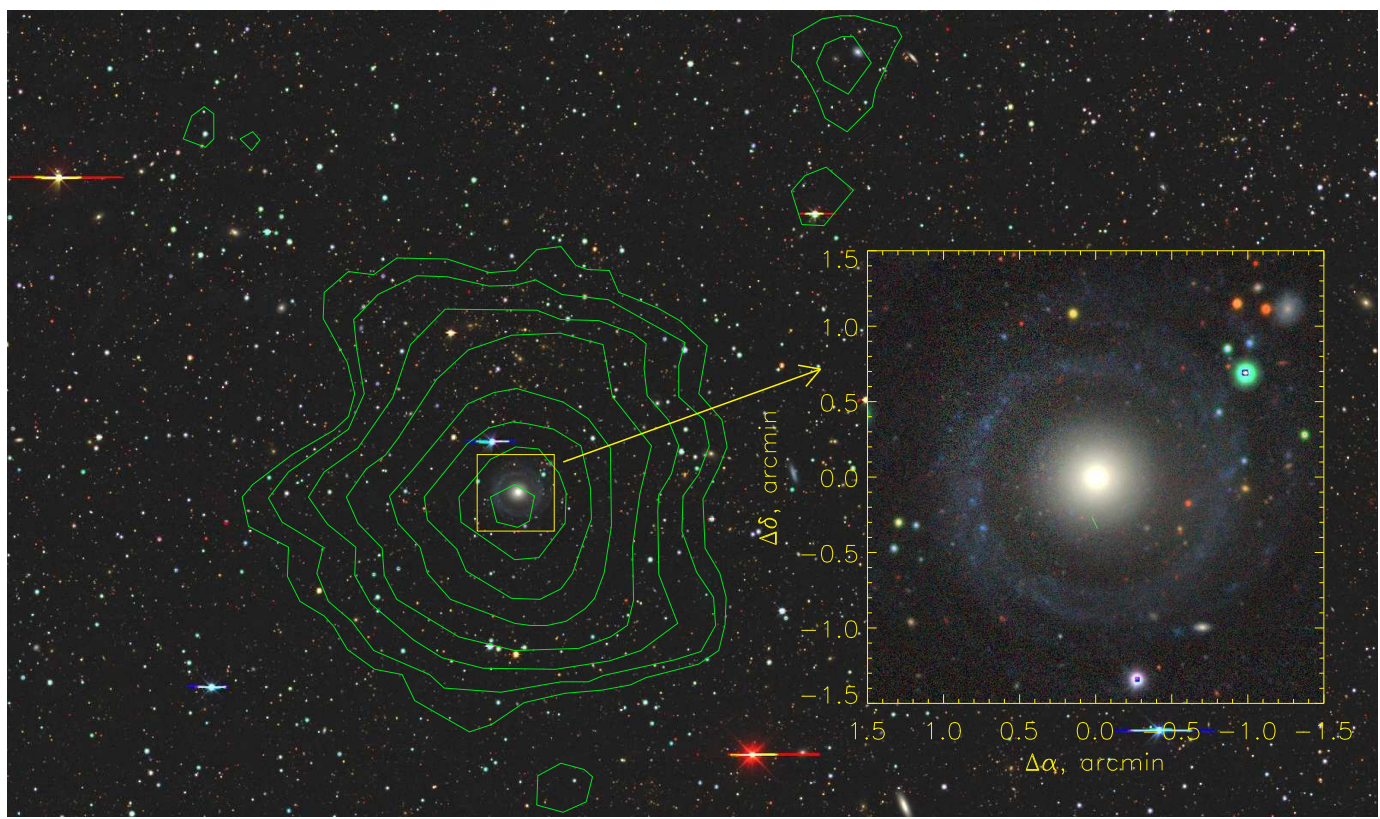}
    \caption{The Legacy Imaging Surveys colour image of the UGC 4599 and its environments. The HI density contours are overlaid following \citet{grossi09}. The north is up, the east is to the left.}
    \label{image} 
\end{figure*}

The galaxy image taken from the LegacySurvey resource \citep{legacysurvey} including the data from the DECaLS photometric survey is presented in Fig.~\ref{image}.
It is seen almost face-on, and with the adopted distance of 32~Mpc (cosmology-corrected luminosity distance from the NASA Extragalactic Database, NED)
the radius of the ring is 7.7~kpc. \citet{hoagtype}
claimed that UGC~4599 is an analog of the famous Hoag galaxy, so being a small elliptical surrounded by a detached ring. 
However there were also other opinions: \citet{dowelldiss} treated UGC~4599 as a classical S0, with the de-Vaucouleurs'
bulge contributing only 32\%\ to the total luminosity, and a low-surface-brightness (LSB) exponential disk. \citet{erwin11} found even more disks:
they classified UGC~4599 as a Type-III-d so discovering {\it two} LSB disks, with the exponential scalelengths of 6.5~kpc
and 10~kpc. Indeed, the deep image in Fig.~\ref{image} allows us to trace the UGC~4599 blue disk extending well beyond the ring radius. 
The galaxy is extremely rich in HI: the data of the ALFALFA survey reveal about $10^{10}$\Ms\ of neutral hydrogen, with the
diameter of the HI disk of $\sim 100$~kpc \citep{grossi09}. We have undertaken further investigation of the galaxy by
applying long-slit spectroscopy, Fabry-Perot interferometry, and narrow-band imaging in the strong emission lines.

\section{Observations and the data involved}

\begin{figure*}[htb!] 
	\centering
	\centerline{
		\includegraphics[height=0.28\textwidth]{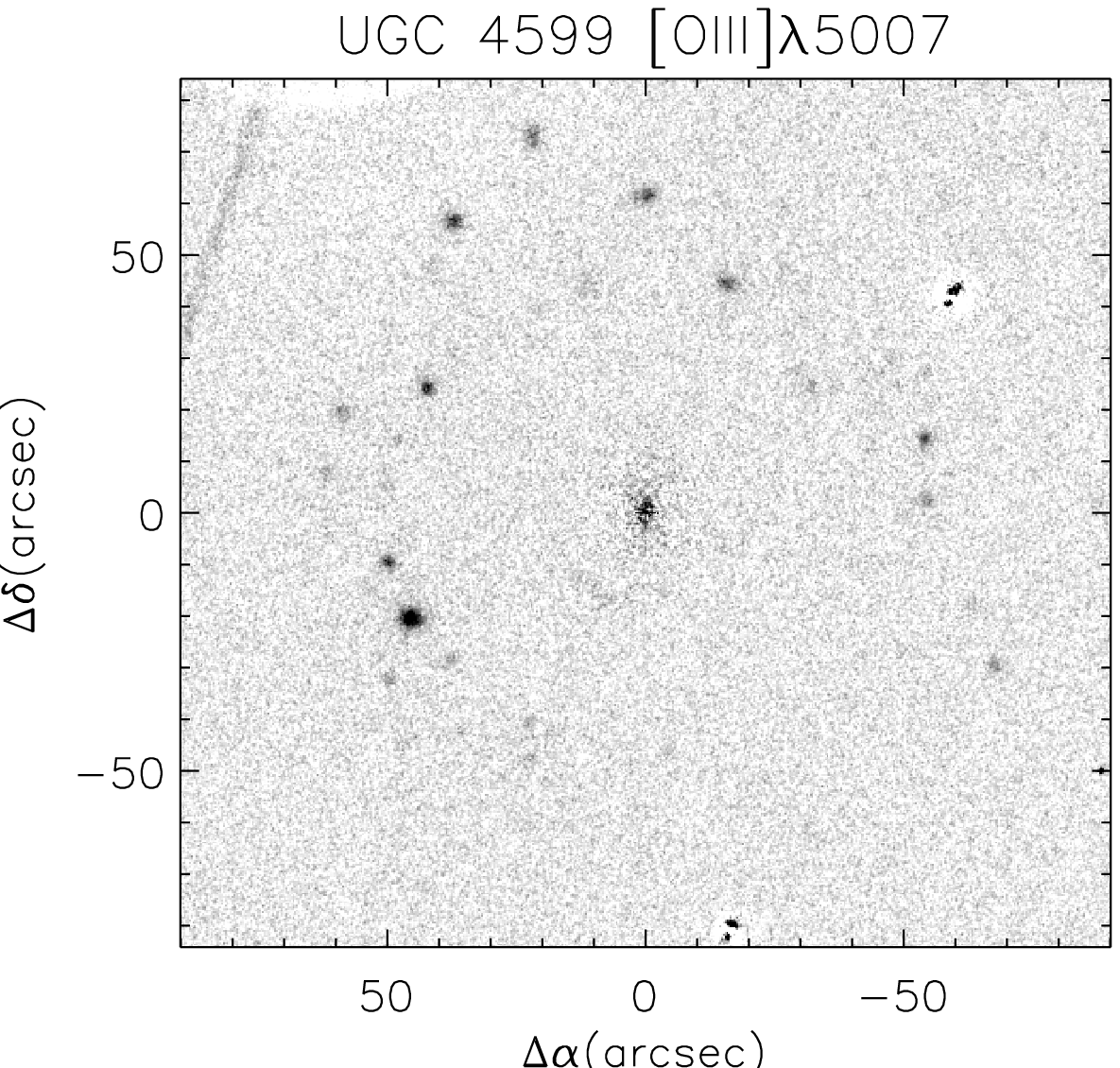}
		\includegraphics[height=0.28\textwidth]{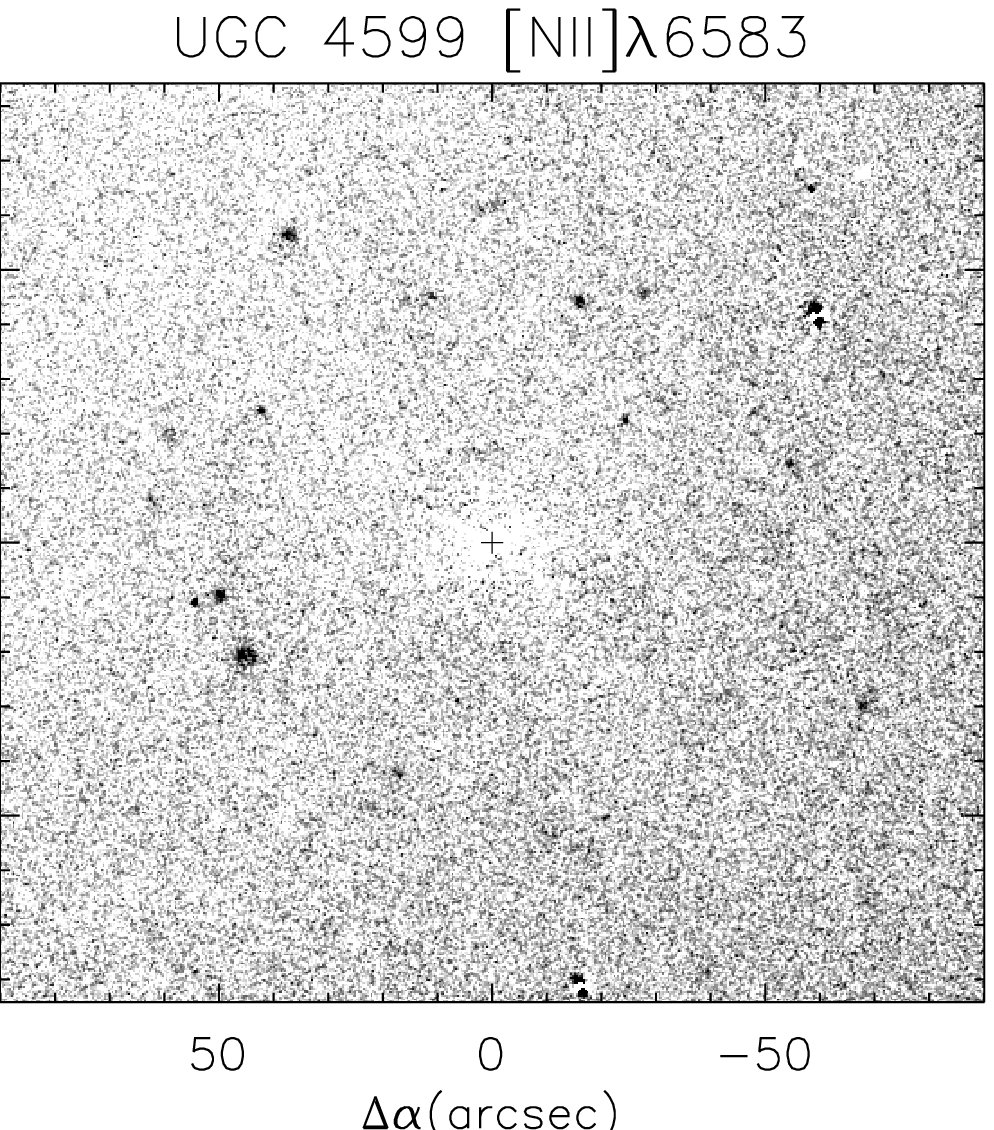}
		\includegraphics[height=0.28\textwidth]{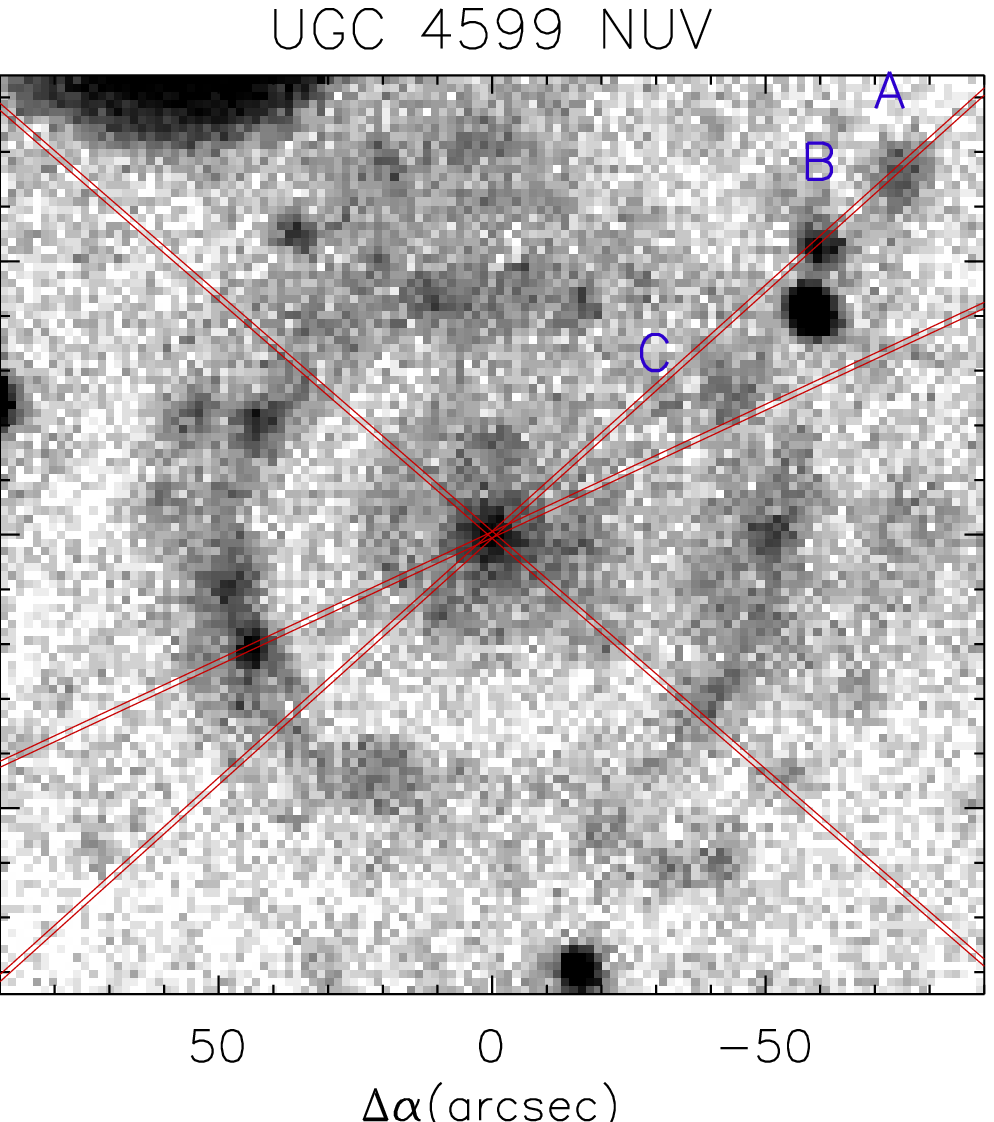}
		\includegraphics[height=0.28\textwidth]{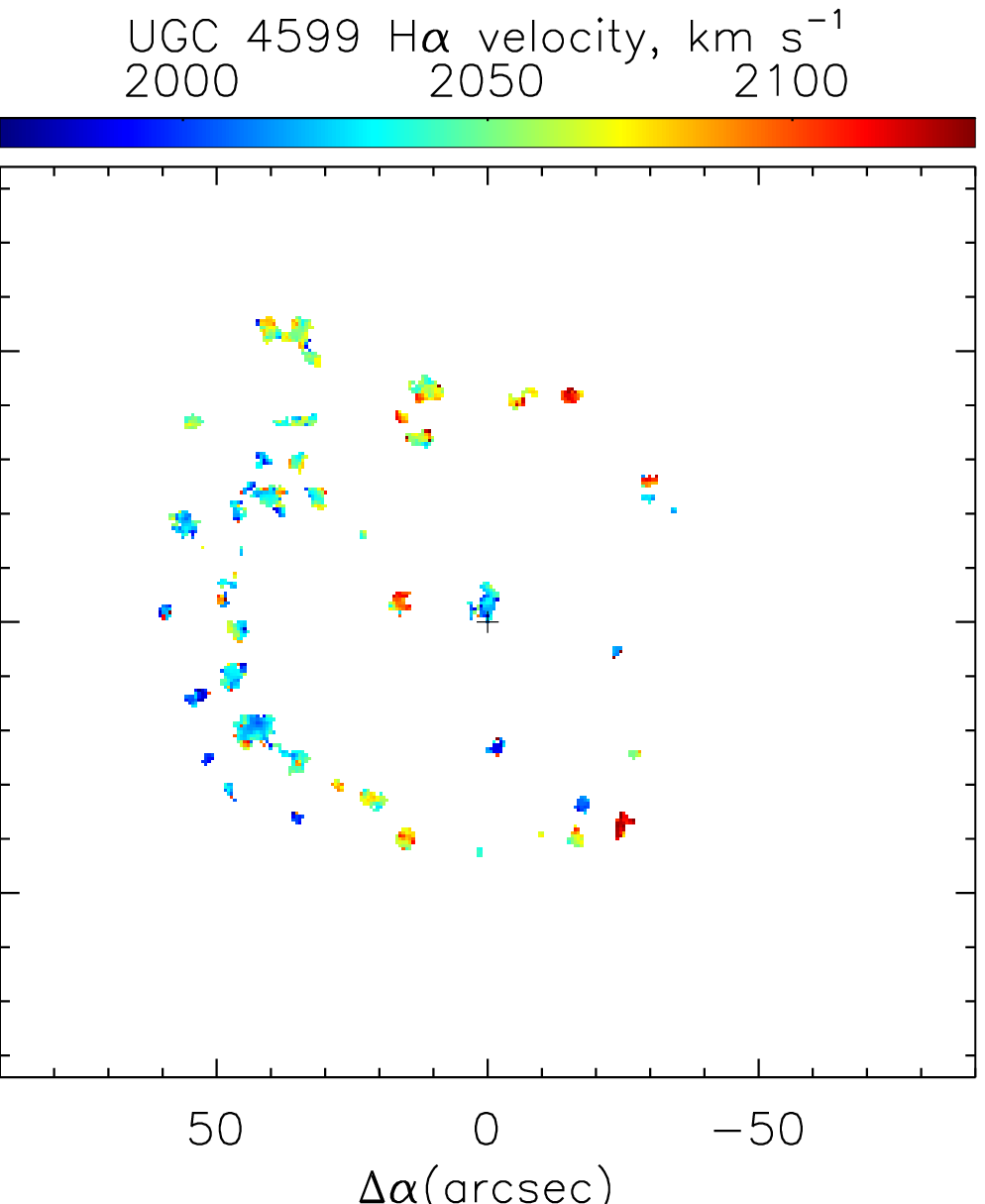}
	}
	\caption{The MaNGaL  images of UGC~4599, from left to right: in the emission line [OIII]$\lambda$5007  and  [NII]$\lambda$6583; then the GALEX NUV image with the SCORPIO-2 slit orientations overplotted and the H$\alpha$ velocity field taken  with GH$\alpha$FaS. At every plot, the north is up, the east is to the left.}
	\label{nuvmangal}
\end{figure*}

Our long-slit spectral observations were made with a multi-mode 
reducer SCORPIO-2 \citep{scorpio2} at the prime focus of the Russian 6m
telescope of the Special Astrophysical Observatory, Russian
Academy of Sciences (SAO RAS). UGC~4599 was observed on February 25th, 2014, at $PA(slit)=115\degr$,
with the total exposure time of 75~min, on March 26th, 2015, in the orientation through
the neighbouring galaxy (designated by {\it A} in Fig.~\ref{nuvmangal}, the third plot) , 
at $PA(slit)=132\degr$, with the total exposure time of 60~min, 
and with a short exposure at $PA(slit)=49\degr$ in February 2022.
The slit orientations are shown in Fig.~\ref{nuvmangal}.
During the observations the range of airmass was 1.2--1.5.
The seeing was $\sim 1''$, 
the VPHG1200 grism provided an intermediate spectral
resolution FWHM $\approx 5$ \AA\ in the wavelength region from 4000~\AA\ to 7200~\AA.
This spectral range includes a set of strong absorption and emission lines
making it suitable to analyse both stellar and gaseous
kinematics of the galaxy as well as the gas excitation
and chemistry.  The slit is 1\arcsec\ in width
and $6'$ in length allowing us to use the edge spectra to subtract the sky
background. The CCD E2V 42-90, with a format of $2048 \times 4600$,
used in the $1\times2$ binning mode provided a spatial scale of 0.357\arcsec\ per px
and a spectral sampling of 0.86~\AA/px. The data reduction as well as the derivation of the characteristics of the gaseous kinematics
were standard for our SCORPIO-2 data -- see for example \citet{n4513_20}. The profiles of the line-of-sight stellar velocity
have been calculated  by  cross-correlation of galactic spectra binned along the slit with  the best-matched template spectra of K-type
stars from the library MILES \citep{miles}.

To study the kinematics in the \Ha\ emission we obtained the data using the Fabry-Perot
Interferometer GH$\alpha$FaS \citep{ghafas} on the William Herschel Telescope (WHT) at the Roque de los Muchachos Observatory, La Palma.
GH$\alpha$FaS has a circular field of view of 3.4 arcmin, free spectral range of 8~\AA\ which
corresponds to 390~\kms\  with a velocity resolution of 8~\kms, with spatial sampling
of 0.2\arcsec. The observations were undertaken on March 14th, 2016, with the total
exposure of 160 s in each of 40 spectral channels. The seeing was 1.2\arcsec.

We also carried out observations at the 2.5m CMO SAI MSU telescope with a narrow-band tuned
photometer MaNGaL \citep{mangal}.
The observations were performed in  two redshifted emission lines, [OIII]$\lambda$5007 and [NII]$\lambda$6583,
within the band of 13~\AA, to study the surface brightness distributions in these lines characterising the ionised gas in the ring.
The detector, CCD iKon-M934 with the format of $1024\times 1024$, provided the field of view of
5.4~arcmin and the sampling of 0.33\arcsec\ per pixel. The observations were undertaken on
November 14th, 2020, for [NII]$\lambda$6583, with an exposure time of 75 min, and on
November 17th, 2020, for [OIII]$\lambda$5007, with an exposure time of 60 min. The seeing
was 1.3\arcsec\ on the first date and 1.9\arcsec\ on the second date.

To study the large-scale structure of the galaxy, we have used the $gr$-band images
from the LegacySurvey (the DECaLS data).

\section{Emission lines in the ring of UGC~4599}

\begin{figure}[htb!]
   \centering
   \includegraphics[width=0.4\textwidth]{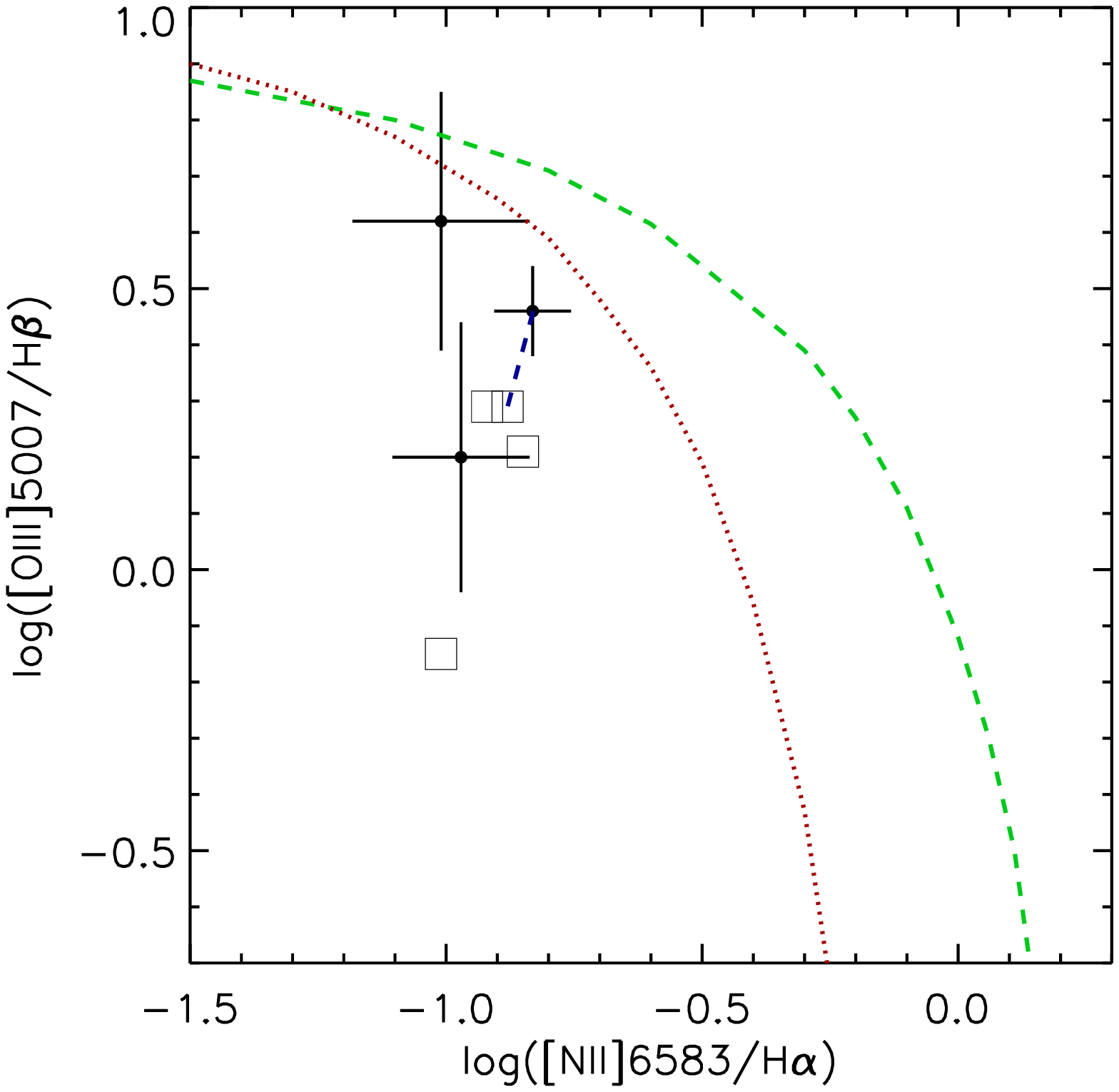}
   \includegraphics[width=0.4\textwidth]{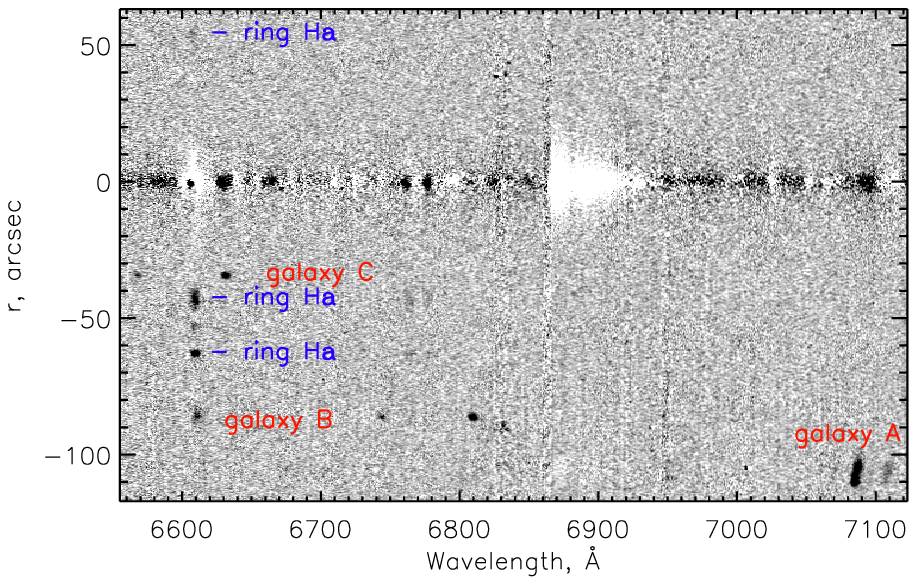}
   \caption{The emission lines in the ring of UGC~4599 and nearby galaxies. In {\it the upper plot} -- the BPT-diagram
for the ring HII-regions, both from our long-slit spectroscopy (with error bars) and from \citet{salzer20} (squares).
The measurements for the common HII-region are connected by a blue dashed line. The excitation-mechanism dividing
lines are from \citet{kewley01} (the green dashed line) and from \citet{kauffmann03} (the red dotted line). In
{\it the bottom plot} -- the view of the red part of the continuum-subtracted long-slit spectrum at $PA=132^{\circ}$.}
\label{bpt_sp}
    \end{figure}

The ring of the galaxy at $R\sim 50$\arcsec\ prominent in the optical continuum (Fig.~\ref{image}) and in the UV  
(Fig.~\ref{nuvmangal}), is also well traced by the emission-line regions in the [OIII]$\lambda$5007 line (Fig.~\ref{nuvmangal}, left):
we can notice more than a dozen compact emission-line sources 
in the Fig.~\ref{nuvmangal} (left). 
Only a few of these regions are also seen in the [NII]$\lambda$6583 emission line (Fig.~\ref{nuvmangal}), and
these detections are mostly concentrated in the eastern half of the ring. This difference in the surface distributions
of the high-excitation and low-excitation emission lines is confirmed by the literature data: \citet{dowelldiss} and \citet{hoagtype} 
analysed the narrow-band deep images of UGC~4599 in \Ha\  obtained within the spectral windows including both \Ha\ and
[NII]$\lambda$6583 and also noted only HII-regions concentrated in the eastern and northern part of the ring.
\citet{salzer20} searched for compact starforming galaxies by using deep narrow-band photometry in the \Ha\ line and
found four HII-regions in the ring of UGC~4599 (their no.103, 104, 105, 106); for these 'HaDots' the full-range spectra
were also obtained at the Hobby-Eberly 9.2m telescope (HET). We use their emission-line ratios together with our results
to inspect the gas excitation (Fig.~\ref{bpt_sp}).

Our long slit at $PA(slit)=115\degr$  crossed the eastern emission-line region, the brightest one in the ring, and our
long slit at $PA(slit)=132\degr$  passed through a small galaxy to the north-west from UGC~4599 which is designated by 'A'
in Fig.~\ref{nuvmangal}. In fact, when we looked at the spectrum we saw a lot of emission-line objects
to the north-west from the centre of UGC~4599 (Fig.\ref{bpt_sp}). Two of them belong to the ring of the galaxy which
is split into two arms to the west; and three of them are background galaxies with  redshifts of 0.08 (A), 0.360 (B), and 0.324 (C).

We use the strong emission-line flux ratios to check the gas excitation by inspecting the so called Baldwin-Phillips-Terlevich(BPT)-diagram \citep{bpt}
in Fig.~\ref{bpt_sp}. All the HII-regions with full-range spectra -- four in the eastern part of the ring measured
by \citet{salzer20} and our data for the brightest eastern region, and two our cross-sections of the ring to the north-west from the centre --
 indeed show HII-type excitation, as they are located to the left of the dividing lines prescribed by \citet{kewley01} and \citet{kauffmann03}.
Then we can determine the ionised-gas oxygen abundance by using the strong-line calibrations. We have involved the  widely used
calibrations from \citet{pm14} and \citet{marino13}. By averaging the oxygen abundances derived from our measurements of N2 and O3N2
for three more bright HII-regions, we have obtained  $12+\log \mbox{(O/H)} =8.23\pm 0.05$~dex from the
\citet{marino13} calibrations. The models by \citet{pm14} involving the measurements of 6 emission lines,
H$\beta$, [OIII]$\lambda$5007, H$\alpha$, [NII]$\lambda$6583, [SII]$\lambda \lambda$6717,6731, give $12+\log \mbox{(O/H)} =8.40\pm 0.11$~dex.

\section{The central stellar spin and the gas kinematics in the ring of UGC~4599}

\begin{figure*}[b]
	\centering
	\includegraphics[width=0.3\textwidth]{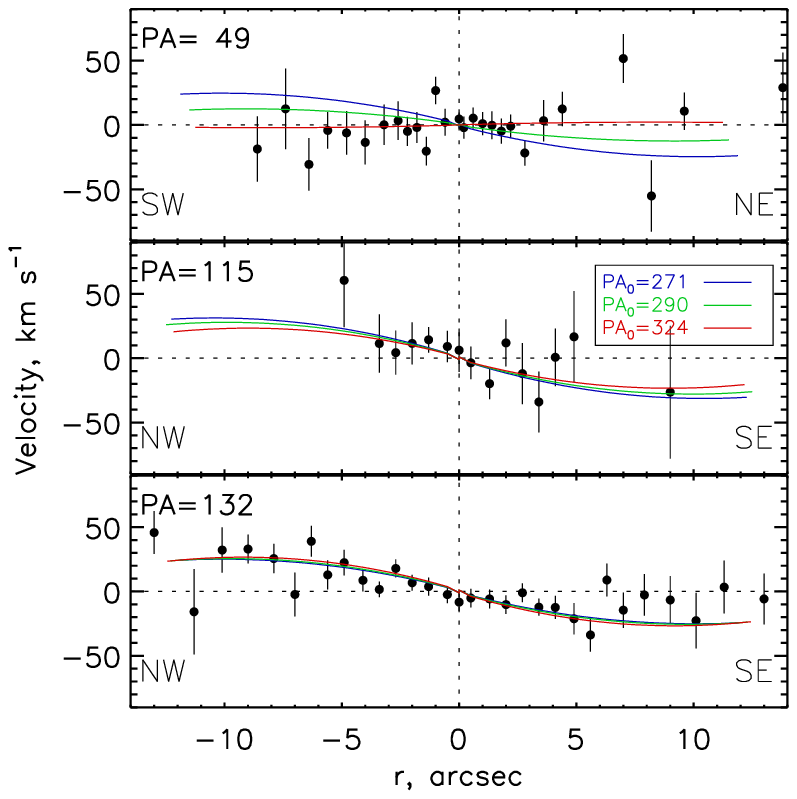}
	\caption{The line-of-sight velocities  of stars along three SCORPIO-2 slit position angles.
        Coloured lines show  the projection of the  second-degree polynomial fitting of the mean rotation
        curve for the various accepted $PA_0$.}
	\label{fig_star}
\end{figure*}

For the main optical emission line of starforming regions, \Ha, we have observed UGC~4599 with the scanning Fabry-Perot
interferometre at the 4.2m WHT to derive not only the \Ha\ map, but also the line-of-sight (LOS) velocity map
reflecting the projection of the ionised-gas rotation in the ring. The results are presented in Fig.~\ref{nuvmangal}, right. Under the
assumption of planar circular gas rotation we can restrict the gaseous disk orientation by using the two-dimensional LOS velocity distribution:
at a fixed radius the projection of the tangential rotation velocity would be maximal at the line of nodes of the gaseous disk
and would be zero in the orthogonal direction. 
Using the tilted-ring model of circular rotation as described in \citet{s0_fp} we obtained the orientation of the line of nodes of the gaseous disk at the radius of the ring: $PA_0=271\degr \pm 9\degr$.
As for the inclination, we can estimate it by plotting UGC~4599, with its HI line width $W_{50}=148$~\kms\ \citep{grossi09} and its baryonic mass, stellar  plus gaseous, $\log M_b =10.18$ \citep{huang14}, onto the baryonic Tully-Fisher relation from e.g. \citet{lelli},
obtaining $i_g \approx 32\degr$. The orientation of the stellar rotation plane can be estimated from the long-slit data. The photometric inclination of the stellar disk was given by \citet{erwin11}
as $i_* = 24\degr$, under the assumption for the disk relative thickness of $q_0=0.2$. Under the more realistic assumption of
the relative typical thickness $q_0=0.4$ found by us for lenticular galaxies in sparse environments \citep{thick_isos0}, this estimate
transforms into $i_* = 26\degr$. The model line-of-sight velocity profiles calculated
with this assumed inclination for three different stellar-disk line-of-nodes orientations are superposed onto our long-slit measurements in Fig.~\ref{fig_star}. It can be seen that the line-of-nodes position angle of the gaseous disk, $PA_0=271\degr \pm 9\degr$, can be excluded for the stellar disk due to zero velocity gradient along the $PA=49\degr$. By fitting all three long-slit cross-sections (Fig.~\ref{fig_star}), the best estimate of the line-of-nodes position angle for the central stellar disk would be $PA_0=324\degr \pm 20\degr$. Evidently, the plane of the gaseous disk is inclined with respect to the central stellar disk.
This configuration is dynamically unstable and gives evidence for recent gas accretion. 

\section{Discussion}

\subsection{The structure and morphological type of UGC~4599}

Opinions on the luminosity and morphological type of UGC~4599 in the literature  show differences.
For example, \citet{hoagtype} treated UGC~4599 as a dwarf {\it elliptical} galaxy, with $M_g=-17.9$ ($M_B=-17.4$)
and a detached 8-kpc ring. However, those  researchers who focused on its extended HI disk also noted  its
large {\it stellar} disk, and then UGC~4599 was classified as an intermediate-luminosity lenticular galaxy:
\citet{grossi09} estimated its integrated absolute magnitude as $M_B=-19.07$ and \citet{dowelldiss} -- as
$M_B=-19.42\pm 0.02$.

\begin{figure*}[htb!]
   \centering
   \vspace{1cm}
   \centerline{
   \includegraphics[width=0.3\textwidth]{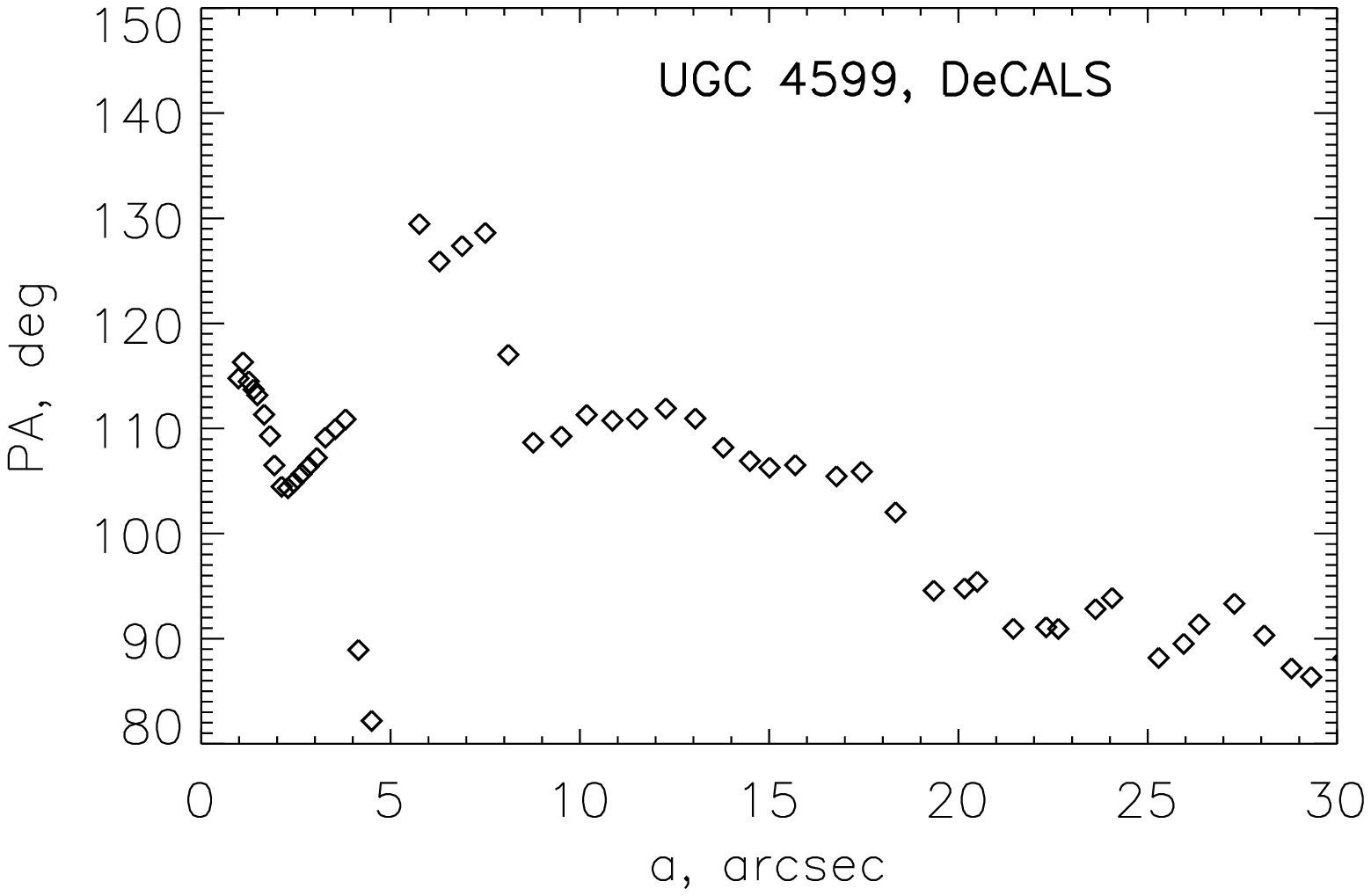}
   \includegraphics[width=0.3\textwidth]{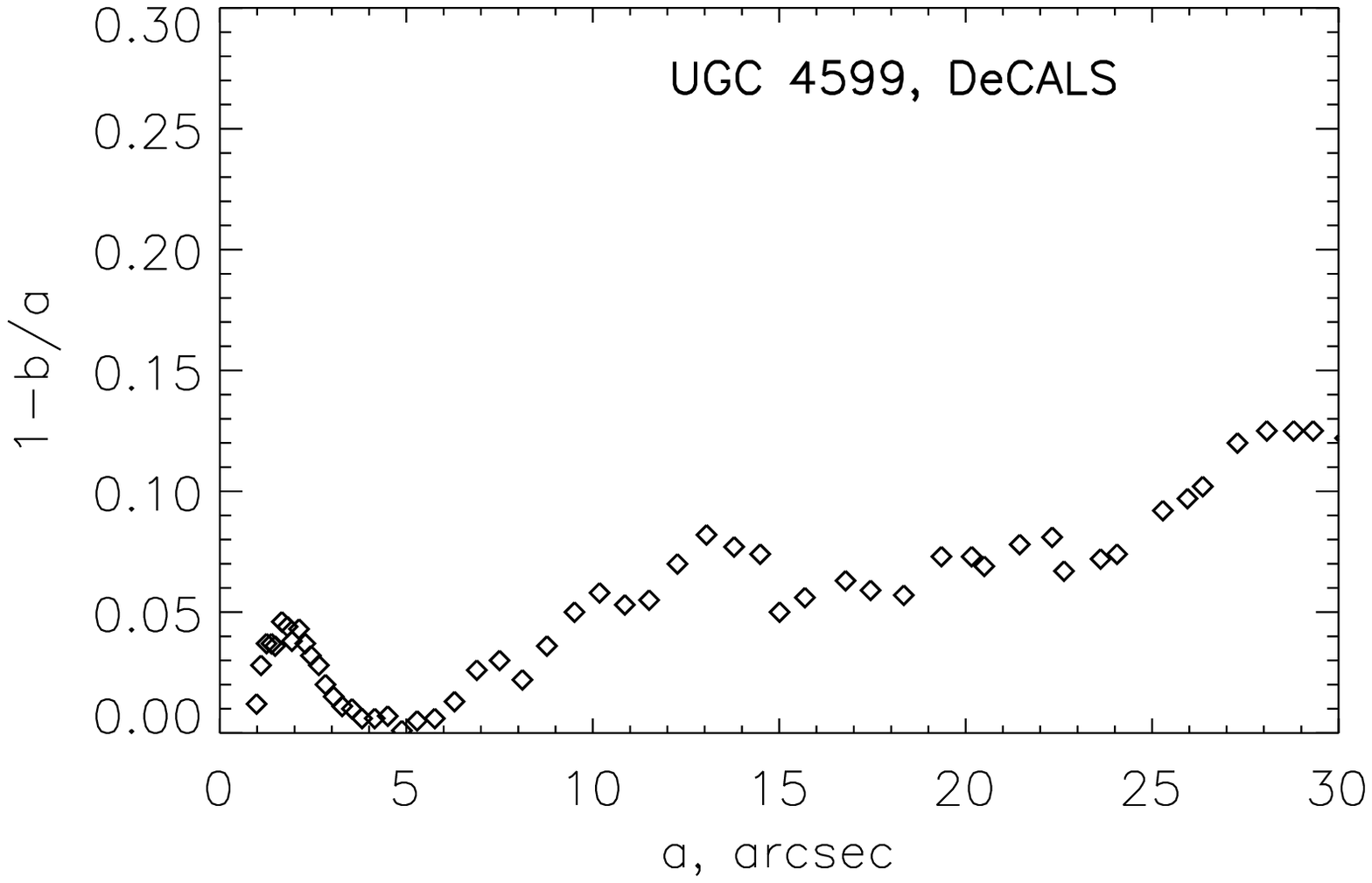}
   \includegraphics[width=0.3\textwidth]{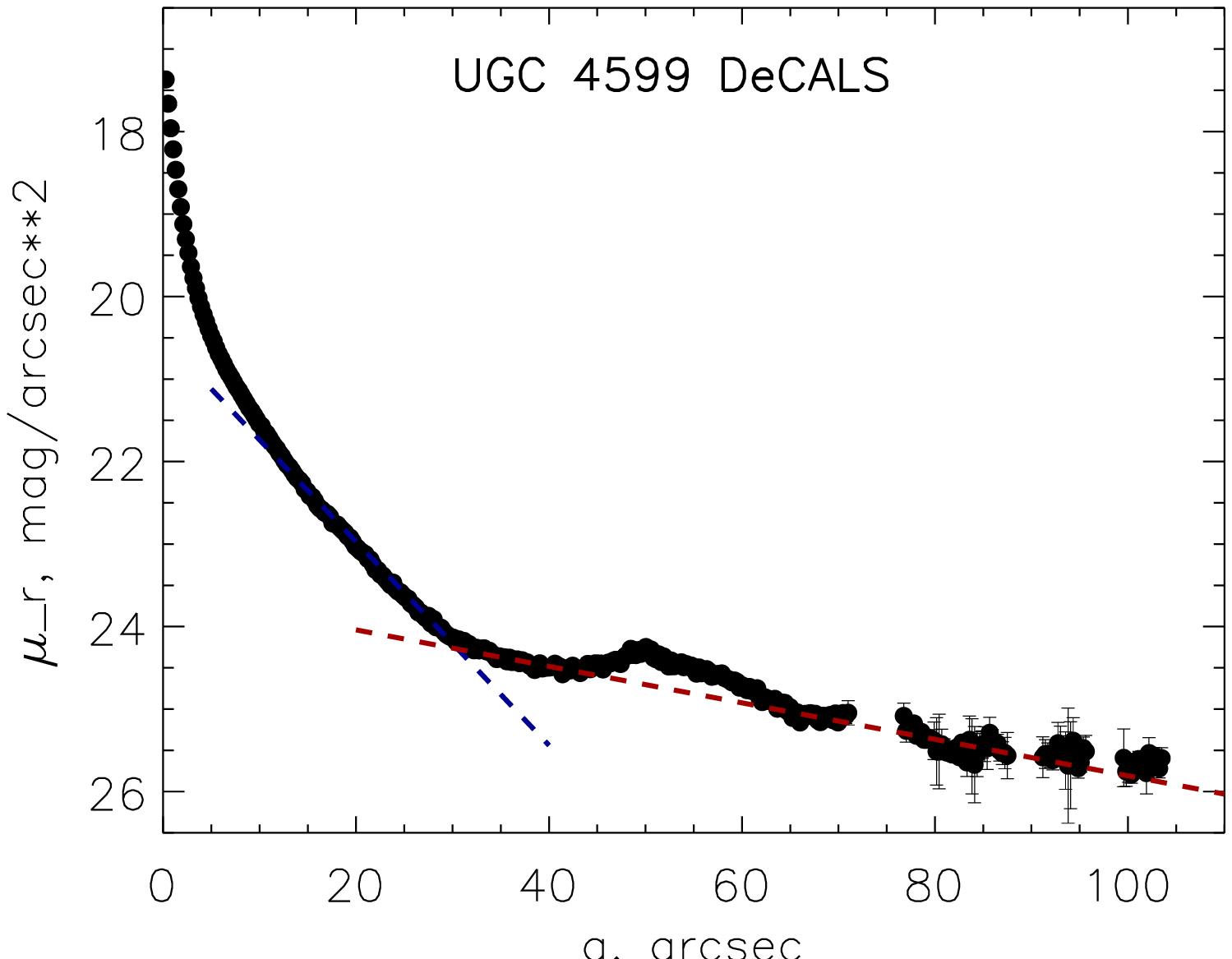}
   }
   \caption{The results of the analysis of the $r$-band UGC~4599 image from the DECaLS survey: isophote characteristics ({\it the left and central plot})
and the azimuthally averaged surface brightness profile with pseudobulge and inner disk fitted by exponential laws ({\it the right plot}).
The two fitted exponential relations are: $\mu _r=20.5 + 1.086 R/8.8^{\prime \prime}$ (blue dashed line) and 
$\mu _r=23.6 + 1.086 R/49.1^{\prime \prime}$ (red dashed line).}
 \label{iso}
    \end{figure*}

We have inspected the rather deep new $gr$-images of UGC~4599 provided by the DECaLS survey.
We present our results of isophote analysis and the azimuthally averaged surface brightness profile of UGC~4599 in Fig.~\ref{iso}.
Indeed, we do not see any gap between the central body (including perhaps the exponential pseudobulge) and the ring:
the ring in the radius range of 45\arcsec --65\arcsec is superposed onto the extended exponential disk which can be
fitted over $R=35\arcsec - 100\arcsec$ range by a model profile $\mu _r=23.6 + 1.086 R/49.1^{\prime \prime}$.
With respect to the measurements by \citet{erwin11}, we have obtained slightly larger exponential scalelength for the
inner portion of the UGC~4599 disk because we have excluded the ring from our fitting. In any case, with its scalelength of
7.5~kpc and its central surface brightness of $\mu _0(r)=23.6$ the galaxy can be classified as an LSB disk galaxy. Moreover,
when we estimate the integrated characteristics of the disk in Fig.~\ref{iso} (right), $M_V=-19.8$ and $r_{eff}=12.6$~kpc,
and compare them with the data in Fig.~12 by \citet{greco} and with the data in Fig.~1 by \citet{glsb}, we ascertain 
that UGC~4599 belongs to the class of {\it giant} LSB disk galaxies and resembles such objects
considered by \citet{glsb} as UGC~1378 or UGC~1382.

\subsection{The origin of the gaseous disk in UGC~4599}

As  was found by \citet{glsb}, the most frequent scenario of giant LSB galaxy formation is accretion of high-momentum gas 
from outside. The second most favoured scenario, coplanar merging of two large spiral galaxies (that was also suggested by \citet{hoagtype}),
can be excluded for UGC~4599 by our data because the low oxygen abundance of the ionised gas in the starforming ring
contradicts the typically solar abundance of gas in non-dwarf spiral galaxies \citep{tremonti_sdss,pilyugin04}. The outer gas accretion scenario is more suitable
also because of the inclined orientation of the gaseous ring spin vector with respect to the collective stellar  angular momentum.
But what can be a source of the outer gas?

The environment of UGC~4599 is rather sparse though the galaxy has been included into galaxy groups by several catalogues: in
USGC~191 by \citet{ramella} and in the compact triplet UZC-CG~79 by \citet{focardi}. Indeed, the group contains three
galaxies of comparable luminosities, UGC~4599, UGC~4590, and UGC~4550. The tightest pair separation, between UGC~4599 and UGC~4590,
is 179~kpc, and the third galaxy, UGC~4550, is classified as completely isolated in the 2MIG catalog \citep{2mig}.
It is important to note that UGC~4590 is devoid of neutral hydrogen though demonstrating a starforming nucleus \citep{grossi09}.
In the latter paper, the authors proposed a hypothesis of gas flow to UGC~4599 from the
nearby dwarf PGC~24666 (CGCG 061-011) because several HI clumps were detected between
UGC~4599 and CGCG~061-011. But CGCG~061-011 is a very small galaxy, with a stellar mass of $3 \cdot 10^8$~\Ms\ \citep{spogs} and HI mass of $4 \cdot 10^8$~\Ms\ \citep{alfalfa};
it seems improbable that CGCG~061-011 may provide a two order larger mass of neutral hydrogen for UGC~4599. Moreover, the metallicity of the ionised gas may be higher
in the dwarf CGCG~061-011 than in UGC~4599. For the latter we have measured 
$\log \mbox{[NII]}\lambda 6583 /\mbox{H}\alpha =-0.95\pm 0.01$ by averaging our emission-line measurements for five HII-regions
(it corresponds to $12+\log \mbox{(O/H)} =8.23$~dex \citep{marino13}).
For CGCG~061-011 we have obtained a long-slit spectrum with the SCORPIO-2 on October 31, 2022. After we have excluded the more metal-rich galaxy core,
the remaining $\log \mbox{[NII]}\lambda 6583 /\mbox{H}\alpha$ profile up to $r=13\arcsec$
has appeared to be flat, and it reveals the mean nitrogen-to-hydrogen line ratio of $-0.809\pm 0.005$ corresponding to
$12+\log \mbox{(O/H)} =8.30\pm 0.09$~dex \citep{marino13}. So the ionised gas in the starforming ring of UGC~4599 is poorer
by oxygen than the gas in the dwarf galaxy CGCG~061-011: it cannot be a donor.
The completed merger of a small gas-rich satellite is not
a good perspective to explain the large HI disk of UGC~4599 too because of its very large mass of accreted neutral hydrogen, $1.1 \cdot 10^{10}$~\Ms\ \citep{alfalfa}; the total stellar mass 
of UGC~4599 is  less than half of this, $M_*=4\cdot 10^9$~\Ms\ \citep{huang14}. 
Therefore we cannot identify a suitable galaxy to play the role of gas donor in the vicinity of UGC~4599.

The metallicity of the ionised gas in the ring of UGC~4599 appears to be unusually low, --0.4~dex. Up to now we have studied
a dozen outer starforming rings in lenticular galaxies, and the gas metallicities in these rings are very homogeneous, $-0.15$~dex
independently of the galaxy luminosity or ring radius \citep{s0_fp,n4513_20}. By confronting the high relative mass of HI and the
low metallicity of the ionised gas, we reach the conclusion that this is perhaps the first clear case of gas accretion to a ring of S0 galaxy
from a cosmological filament. Then the chain of HI clumps  in the direction from UGC~4599 to PGC~24666 \citep{grossi09} may trace this filament.
A quite similar structure has been observed in 21~cm near the Hoag object which is a recognised case of filamentary gas accretion onto an elliptical
galaxy \citep{hi_hoag}.

\begin{acknowledgements}
This study is based on the data obtained at the   unique scientific facility   the Big Telescope Alt-azimuthal  SAO RAS and  was supported  under  the   Ministry of Science and Higher Education of the Russian Federation grant  075-15-2022-262 (13.MNPMU.21.0003).
The renovation of 6m telescope equipment is currently provided within the national project ''Science''. The work used  the public data of the
Legacy Surveys (http://legacysurvey.org), that  consists of three individual and complementary projects: the Dark Energy Camera Legacy Survey (DECaLS; Proposal ID $\sharp 2014B-0404$;
PIs: David Schlegel and Arjun Dey), the Beijing-Arizona Sky Survey (BASS; NOAO Prop. ID $\sharp 2015A-0801$; PIs: Zhou Xu and Xiaohui Fan), and
the Mayall z-band Legacy Survey (MzLS; Prop. ID $\sharp 2016A-0453$; PI: Arjun Dey). DECaLS, BASS and MzLS together include data obtained,
respectively, at the Blanco telescope, Cerro Tololo Inter-American Observatory, NSF’s NOIRLab; the Bok telescope, Steward Observatory,
University of Arizona; and the Mayall telescope, Kitt Peak National Observatory, NOIRLab. The Legacy Surveys project is honored to be permitted
to conduct astronomical research on Iolkam Du\'ag (Kitt Peak), a mountain with particular significance to the Tohono O\'odham Nation.
The NASA GALEX mission data for our Fig.~2 were taken from the Mikulski Archive for Space Telescopes (MAST).
The William Herschel Telescope is in the Isaac Newton Group of telescopes, situated at the Roque de los Muchachos Observatory, La Palma, of the Instituto de Astrofisica de Canarias (IAC).
\end{acknowledgements}

\end{document}